# Real-time Monitoring and Early Warning Analysis of Urban Railway Operation Based on Multi-parameter Vital Signs of Subway Drivers in Plateau Environment

Zhiqiang Sun[1]，Chaozhe Jiang[2,3]*, Yongjie Lu[1], Chao Wen[2,3]
Xiaozuo Yu[4], Tesfaya Hailemariam Yimer[2]

1, Lanzhou Rail Transit Co., Ltd
2, School of Transportation and Logistics, Southwest Jiaotong University, Chengdu, China
3, Department of Civil and Environmental Engineering, University of Waterloo, Canada
4, Amazon Toronto, Toronto, Canada

**Email:**
wps3000@163.com
*jiangchaozhe@swjtu.edu.cn (corresponding author)
lyj18919075246@163.com
wenchao@swjtu.cn
jamesonyu95@gmail.com
tsfhailemariam04@my.swjtu.edu.cn

**Abstract**：In order to ensure the personal safety of the drivers and passengers of rail transit in plateau environment, the vital signs and train conditions of the drivers and passengers are taken as the research object, and the dynamic relationship between them is studied and analyzed. In this paper, subway drivers under normal operation conditions are taken as research objects to establish the vital signs monitoring and early warning system. The vital signs data of the subway drivers, such as heart rate (HR), respiratory rate (RR), body temperature (T) and blood oxygen saturation ($SPO_2$) of the subway driver are collected by the head-mounted sensor, and the least mean square adaptive filtering algorithm is used to preprocess the data and eliminate the interference information. Based on the improved BP (Back Propagation) neural network algorithm, a prediction model is established to predict the vital signs of subway drivers in real-time. We use the early warning score evaluation method to measure the risk of subway drivers' vital signs, and then the necessary judgment basis can be provided to dispatchers in the control center. Experiments show that the system developed in this paper can accurately predict the evolution of subway drivers' vital signs, and timely warn the abnormal states. The predicted value of vital signs is consistent with the actual value, and the absolute error of prediction is less than 0.5 which is within the allowable range.

## 1.Introduction

In recent years, with the rapid development of urban rail transit (especially in China), the need to ensure personal safety of drivers and passengers has become a key issue. Generally speaking, monitoring the vital signs of the drivers and passengers can reflect the physical condition of the human body (normal or abnormal), but there are various factors that affect the changes of vital signs, such as the operation conditions of the train, environmental conditions and physical signs of the train itself. Therefore, a single judgment index cannot meet the applicability requirements under complex conditions.

The operation conditions of metro trains mainly include constant speed operation, acceleration, deceleration, braking and emergency braking. At the same time, due to the high altitude, the plateau environment has unique geographical conditions and natural environments, such as low air pressure, low oxygen partial pressure, cold and strong wind, low humidity, strong solar radiation, etc., which will have an impact on both the human physiology and psychology. These impacts will be reflected in the vital signs data of the drivers and passengers. Therefore, this paper will study and explore the dynamic mechanism of real-time vital signs and train conditions of rail transit drivers and passengers in plateau environments.

Lanzhou Metro is in the Loess Plateau of China. It is a typical plateau environment, and its environmental conditions are shown in Table 1.

Table 1 Introduction to the environment of Lanzhou Metro

| Factor | Values |
| --- | --- |
| Highest altitude | 2171m |
| Lowest altitude | 1503m |
| The annual average temperature | 10.3°C |
| Average annual precipitation | 327mm |
| Annual average sunshine hours | 2446h |

Therefore，subway drivers' performance is one of the most crucial elements related to train operation such an environment. Furthermore, subway drivers often work in a closed, monotonous, and light deficient environment, which leads to more severe status quo. To ensure that subway drivers maintain a good psychological and physical conditions, subway operators generally use psychological training[1], questionnaire survey[2], alcohol testing instrument test[3] and other methods to test the psychological and physical conditions of subway drivers. However, the above-mentioned methods can only appropriately relieve the working pressure of subway drivers and judge the psychological quality of drivers subjectively, which cannot find the potential risks due to increasing workload and time span. In addition, the current methods can only judge the mental state and physical condition of subway drivers statically before they go on duty and cannot monitor and measure the real-time states when on duty.

At present, driverless technology[4] is also applied in the field of urban rail transit, but it is only applied to the part of automatic passenger rapid transit system[5], tram[6] and metro line[7] with small passenger flow, and a certain number of maintenance personnel are allocated during the operation of electric bus, which does not realize the true sense of driverless. In addition, unmanned driving also faces many challenges[5-7], such as higher requirements for reliability of the signal system and dispatching management system, higher requirements for system stability and safety in a complex environment, higher requirements for emergency response of drivers in emergency situations, higher requirements for maintenance of facilities and equipment.

It is of great significance and practical value to ensure the life safety of passengers and the safe operation of the subway through real-time monitoring of vital signs of subway drivers. Therefore, this paper establishes the monitoring and early warning system of vital signs of subway drivers in plateau environment under the condition of constant train speed, to ensure the personal safety of drivers and passengers.

First, we preprocess the collected data to eliminate noise signals. Then, we use the improved BP neural network algorithm to establish the mathematical model of vital signs prediction and carry out the experimental verification of the model. Finally, we evaluate the risk of abnormal vital signs of subway drivers through the early warning score to classify conditions.

The main contribution of this paper is to use our developed system to monitor the vital signs of subway drivers on duty in real time without human interventions, and provide the necessary judgment basis for the control center dispatchers to make relevant decisions. At present, there does not exist a set of related system applied to the real-time monitoring of subway drivers' vital signs thus making the research work of this paper very innovative. The system we developed will bring great practical value in ensuring the safe operation of subways.

The rest of this paper is arranged as follows. Section 2 analyzes the research status of vital signs detection technology. Section 3 outlines the functions and principles of the system. Section 4 outlines the prediction model for vital signs. Section 5 describes the risk assessment of vital signs. Section 6 describes the experimental tests and results. Finally, section 6 is the conclusion of this paper.

**2. Literature review**

Through the monitoring of vital signs such as HR, RR, T, $SPO_2$, we can judge the physical condition of subway drivers [8-11]. It has become an important research topic in recent years to judge the physical condition of the human body by detecting the above-mentioned vital signs. The detection methods of vital signs mainly include contact detection and non-contact detection, which are widely used in post-disaster search and rescue[12], through-wall monitoring[13], medical diagnosis and monitoring[14] and family health monitoring[15]. The commonly used non-contact vital signs detection methods include optical vital signs detection[16,23-24], audio vital signs detection[17], infrared vital signs detection[18] and radar vital signs detection[10,19-20]. Contact vital signs detection method monitors all vital signs of the human body through wearable detection equipment.

There are many vital signs detection technologies in many literatures, but most of the researches focus on non-contact vital signs monitoring and analysis [8-21]. This paper describes the technology of contact vital signs monitoring [22-26]. In [22], a noninvasive vital sign monitoring method based on the change of contact capacitance between passive sensor and human body is designed. Different from the traditional capacitive coupling, the electrical isolation capacitance sensor is used to monitor the heart rate and remove the motion artifacts by processing the signals collected from multiple contact points. In [23], the gradient coil integrated on the driver's seat is used for non-contact vital signs monitoring of the driver. When the driver is in the static stage such as autonomous driving or road driving, the sensor can measure the respiratory activity, and recognize the heartbeat when there is no respiratory activity, but the signal-to-noise ratio is low, so further signal processing is needed to extract the signal. In [24], a new method of using non-invasive optical fiber sensor to identify respiratory events is proposed. After the optical fiber signal is demodulated, the breath is judged by observing the change of heartbeat amplitude. In [25], a multi-method heartbeat and respiration detection method based on optical interference signal is designed. When the optical interferometer directly or indirectly contacts with the human body, the mechanical and acoustic activities of the heart muscle and respiration are reflected in the interference signal, so that the monitoring of heartbeat and respiration is completely free from interference. This method consists of two stages: the first stage is to select the best detection method for a group of signals, the second stage is to test a group of signals with the selected method, and fuse all the detected vital signs.

From the above research, it can be found that the monitoring of vital signs mainly focuses on the monitoring and analysis of relatively single vital sign parameters, which cannot meet the requirements of the special working environment for subway drivers. To monitor the vital signs of subway drivers more comprehensively, the system adopts contact wearable equipment to collect numerous vital signs parameters of subway drivers. Through the improved BP neural network algorithm [27-30] to process and analyze the collected data, which can realize the real-time monitoring and early warning of the subway driver's physical condition.

## 3.Monitoring and early warning system

The goal of this paper is to develop a system with high real-time and accuracy. It can not only monitor the vital signs of subway drivers in real time, but also predict the vital signs in real time and give early warning of abnormal conditions. Because the system has accurate prior prediction ability, it can provide an effective management tool for the safe operation of subway. The system structure is shown in Figure 1.

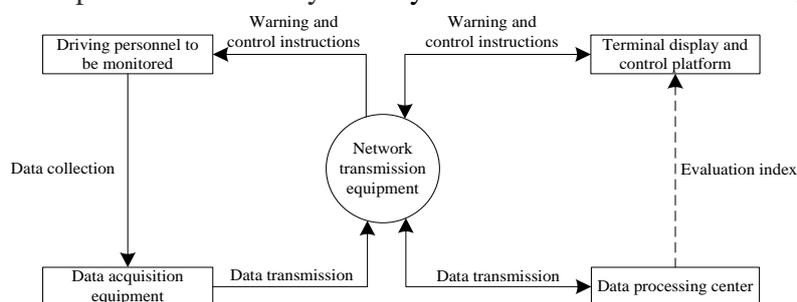

Fig1 System structure

The challenge here is to monitor multiple vital signs of subway drivers in real time without any human intervention. This paper does not put forward any new data prediction method, but through the combination of adaptive filtering algorithm, improved BP neural network algorithm and early warning model to effectively predict the vital signs of subway drivers and realize early warning of abnormal conditions. The whole process includes the following steps.

(1) Sensors are used to collect multiple vital signs data from subway drivers.

(2) Filtering algorithm is used to preprocess the collected vital sign information and eliminate the interference signals.

(3) The vital signs of subway drivers are selected as training samples and test samples, respectively.

(4) The prediction model is established by combining genetic algorithm and BP neural network algorithm.

(5) The prediction performance of the model is verified by the application of the model to the prediction of vital signs of randomly selected subway drivers.

(6) The true positive rate and false positive rate are used to evaluate the prediction results.

(7) The risk of abnormal vital signs is evaluated by early warning score.

The collection of vital signs data is realized by wearable sensors with wireless communication ability, which can collect a variety of vital signs data at the same time. Most of the collected vital signs data are stored in the database of the control center. The least mean square adaptive filtering algorithm is then used to eliminate the interference signals generated by the complex environment in the collected signals, to provide vital signs data that meet the prediction requirements.

The improved BP neural network algorithm is used for real-time prediction of subway drivers' vital signs. The trained model is stored in the repository located in the control center for data prediction in the future. Then test samples are used to verify the training results. If the system can predict the occurrence of abnormal vital signs, it will notify the corresponding monitoring service. The dispatcher of the control center can make necessary adjustments according to the physical condition of the subway driver and give appropriate warning to the subway driver.

The evaluation of the learning model is based on the accuracy, training time and prediction time, while the performance of the real-time prediction system is evaluated by the true positive (TP) rate and false positive (FP) rate of each subway driver's vital signs prediction value.

The prediction results were evaluated by early warning scores to judge the severity of abnormal vital signs of subway drivers. According to the numerical range of the prediction results, the corresponding warning scores are given, i.e. 0, 1, 2 and 3. Through the early warning score evaluation, it can provide the judgment basis for the dispatcher to make the corresponding decision.

Four kinds of biological signals of subway drivers, namely HR, RR, T and $SPO_2$ are selected for analysis. These data are continuous values, collected every 3 seconds. Here, normality is defined using the general normal range of vital signs in the medical rules. When one or more vital signs of metro drivers are detected without the expected normal range, they are defined as abnormal conditions. In this paper, the abnormal situation is divided into general situation and emergency situations. When the abnormal situation lasts for more than a fixed time *t*, it is an emergency. Otherwise, it is a general situation. If the system cannot predict the abnormal situation before it happens, the dispatcher cannot act in time. Therefore, the prediction horizon period selected in this paper is at least 30 minutes, which provides enough time for the dispatcher to make correct decisions.

**4.Prediction model**

4.1Data collection and preprocessing

The subway driver is basically in motion during his/her work. The vibration, noise and other signals generated by the vehicle in the operation process will interfere with the detected vital signs data. Because the system uses wearable contact equipment to detect the vital signs of subway drivers, relative movements between the probe of signal detection equipment and the tested part may occur during the movement, and interference may be introduced. Additionally, the frequency band of the interference signal generated by the environment noises and motion overlaps the frequency band of the target signal, which will directly affect the accuracy of the detection results. Traditional frequency filtering method cannot meet our requirements and therefore, to realize the real-time and accurate prediction of subway drivers' vital signs, it is necessary to filter the collected vital signs effectively and eliminate the interference signals. The signals collected by the wearable vital signs' detection equipment are usually the pulse wave signals of the optical capacitance product. Due to the characteristics of the interference signals of the objects studied in this paper, the least mean square adaptive filtering algorithm is adopted to deal with the interference signals generated by environmental noise and motion.

The input signal, output signal, error value and reference signal of the minimum mean square adaptive filter are defined as $x(k)$, $y(k)$, $e(k)$, $d(k)$, and the output signal, error value and update equation can be expressed as:

$$y(k) = w^\mathrm{T}(k) * x(k) = \sum_{i=0}^{N-1} w_i(k) x(k-i) \tag{1}$$

$$e(k) = d(k) - y(k) = d(k) - w^\mathrm{T}(k) * x(k) \tag{2}$$

$$w(k+1) = w(k) + \mu e(k) * x(k) \tag{3}$$

Where, $T$ represents transposed matrix, $N$ represents filtering order, $\mu$ represents convergence factor, and the range is $(0, \frac{1}{\lambda_{\max}})$, $\lambda_{\max}$ is the maximum eigenvalue of matrix R.

The basic work steps of the least mean square adaptive filter are:

(1) Given initial weight parameter of filter $w(0)$.

(2) Calculate the output value $y(k)$.

(3) Calculate the error value $e(k)$ of the $k$ time from the value of $w(k)$, $y(k)$, and $d(k)$.

(4) Use formula (3) to update the weight parameters of the filter.

(5) Noting that $k = k+1$, and repeat the above steps (2) - (4) to reach a stable state.

To verify the superiority of the filtering method used in the system in the moving environment, the actual collected heart rate data of subway drivers are randomly selected and converted into voltage signals for filtering. The results are shown in Figure 2.

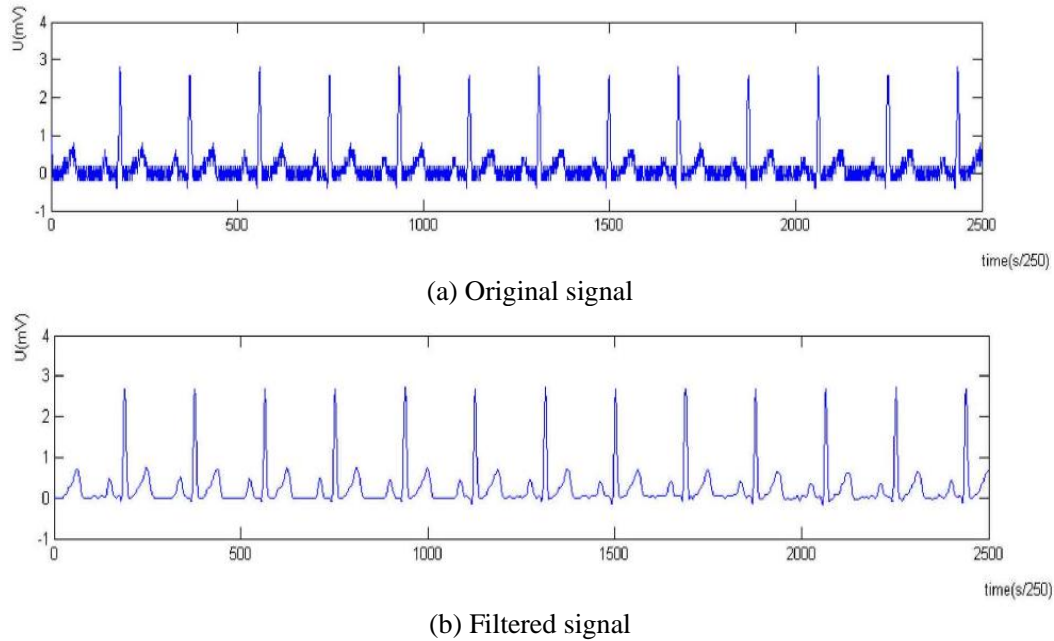

(a) Original signal

(b) Filtered signal

Fig2 Data preprocess

It can be seen from Figure 2 that the filtering algorithm adopted in this system can well eliminate all kinds of interference signals contained in the collected original signals, and can provide the required data for the prediction of vital signs of subway drivers.

4.2 Establishing prediction model

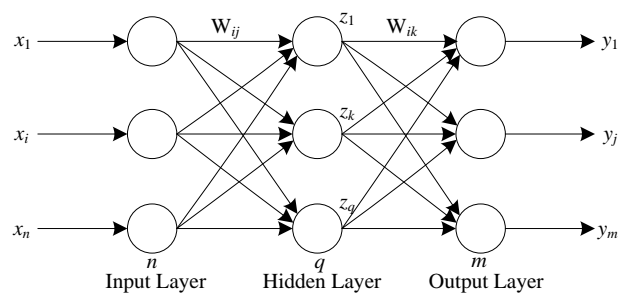

Fig3 Structure graph of 3 layered BP network

BP neural network trains and learns through the back propagation of error and the forward propagation of signal, and constantly correcting the weights and thresholds between each neuron until the output value and expected value meet the error requirements. After experimental verification, this paper adopts a three-layer BP neural network structure, as shown in Figure 3.

Although BP neural network has good generalization ability and self-adaptive ability, the initial weights and thresholds of the network are randomly selected, which will fall into the local minimum when finding the global optimal solution. Therefore, this paper uses the method of combining genetic algorithm and BP neural network algorithm to search the initial weight and threshold population globally, to get the most suitable optimal individual to assign value to the network, and then uses the local optimization ability of BP algorithm to get the optimal solution. The specific steps are as follows.

(1) Population initialization. Chromosome in genetic algorithm is composed of weight and threshold of each layer in BP neural network, and coding is carried out.

(2) Calculate the fitness value $F$.

$$F = k\left[\sum_{i=1}^{n}(y_i - o_i)^2\right] \quad (4)$$

Where, $k$ is the coefficient, $n$ is the number of network output nodes, $y_i$ is the expected output value of the *ith* node of the neural network, $o_i$ is the actual output value of the *ith* node.

(3) Determine the selection probability of each individual.

$$f_i = \frac{k}{F_i} \quad (5)$$

$$P_i = \frac{f_i}{\sum_{j=1}^{T} f_j} \quad (6)$$

Where, $T$ is the number of individuals in the population, $f_i$ is the fitness value of individual $i$.

(4) Chromosome crossing.

$$\begin{aligned} b_{kj} &= b_{lj}a + b_{kj}(1-a) \\ b_{lj} &= b_{kj}a + b_{lj}(1-a) \end{aligned} \quad (7)$$

(5) Chromosome variation.

$$b_{ij} = \begin{cases} b_{ij} + f(g)(b_{ij} - b_{max}) & r > 0.5 \\ b_{ij} + f(g)(b_{min} - b_{ij}) & r \leq 0.5 \end{cases} \quad (8)$$

$$f(g) = r'(1 - \frac{g}{G_{max}}) \quad (9)$$

Where, $b_{ij}$ represents the *jth* gene of the *ith* individual in the range of $[b_{min}, b_{max}]$. $r$ is a random value in the range $[0,1]$. $r'$ is a random value. $g$ is the number of iterations. $G_{max}$ is the maximum number of evolutions.

(6) The optimal chromosome is calculated as the initialization weight and threshold of BP network.

(7) Use the improved BP neural network model to predict vital signs.

After a large number of experiments, the relevant parameters are set as follows: $T = 50$, $G_{max} = 80$, the probability of crossover is 0.5 and the probability of variation is 0.06.

After the establishment of the prediction model, the vital signs of subway drivers are used as the input values for data prediction.

**5. Early warning score evaluation**

For the predicted vital signs data of subway drivers, this paper uses the method of early warning score evaluation to further analyze and provide the necessary judgment basis for the control center dispatcher to make the corresponding decision.

From the medical point of view, vital signs are used to judge the severity and criticality of the patient's condition, mainly including the changes of heart rate, pulse, blood pressure, respiration, pain, blood oxygen, pupil and corneal reflex. Considering the actual situation of subway drivers, this paper selects HR, RR, T and $SPO_2$ as the main indicators of subway drivers' physical condition, and the normal value range [14] is shown in Table 2. Here, it is particularly noted that the temperature is collected from the forehead of the subway driver.

Table 2 Normal thresholds of different variables

| Variables | Normal thresholds |
| --- | --- |
| HR | 60-100Times/Minute |
| RR | 12-17Times/Minute |
| T | 36.0°C-37.3°C |
| $SPO_2$ | 93-100% |

There are two ways to establish the warning score: manual data recording and continuous data collection [31-32]. In this paper, the data of the system is collected by wearable sensor in 3 seconds, which is a continuous time series data. Therefore, this paper uses the early warning score based on continuous data collection to evaluate the risk of abnormal vital signs of subway drivers. The warning scores of HR, RR, T and SPO$_2$ are shown in Table 3.

Table 3 Warning scores of different variables

| Variables | Scores | | | | | | |
|---|---|---|---|---|---|---|---|
| | 3 | 2 | 1 | 0 | 1 | 2 | 3 |
| HR | ≤50 | 51-58 | 59-63 | 64-104 | 105-112 | 113-127 | ≥128 |
| RR | ≤7 | 8-10 | 11-13 | 14-25 | 26-28 | 29-33 | ≥34 |
| T | ≤35.4 | - | 35.5-35.9 | 36.0-37.3 | 37.4-38.3 | - | ≥38.4 |
| SPO$_2$ | ≤84 | 85-90 | 91-93 | ≥94 | - | - | - |

To establish a score evaluation mechanism for the early warning of vital signs of subway drivers and evaluate the risk of abnormal vital signs, which can not only realize the real-time monitoring of the physical condition of the subway drivers, but also provide a certain management basis for the safe operation of the subway.

**6.Experimental test and performance evaluation**

To evaluate the performance of the multi parameter real-time monitoring and early warning system for subway drivers, and to illustrate the practical application value of the system in the risk analysis of the physical condition of subway drivers on duty, this paper tests the developed system based on a large number of actual collected vital signs data.

6.1Data preparation

In the experiment evaluation of this paper, the database based on the electric bus driver of Lanzhou Metro is used. The database contains 187 subway drivers' vital sign data such as HR, RR, T and SPO$_2$, which are all continuous time series data collected by wearable sensors. Based on the analysis and statistics of 187 drivers, it is found that there are 15 drivers with Metro driving experience, 36 drivers with railway driving experience and 136 drivers without driving experience. The vital sign data of subway drivers from different sources have different degrees of difference, so it is necessary to verify the applicability of our system to different monitoring objects.

From November 2018 to January 2020, the vital signs of 187 Metro drivers were monitored. These data run through the stages of commissioning, trial operation, and formal operation. Therefore, this paper has many comprehensive data to evaluate the performance of the developed system. The database is fully compliant with the criteria for the development and evaluation of monitoring and early warning systems.

Because of the influence of complex environment, the database inevitably contains noise data. Therefore, the data preprocessing algorithm mentioned above is used to eliminate the noise signal and provide a database that meets the requirements for the prediction of vital signs.

6.2 Test results

The goal of this paper is to effectively predict the abnormal vital signs in most cases by learning less times. Through experiments, the performance of the developed prediction model in prediction accuracy and efficiency is verified. To achieve the purpose of fast learning, the industrial computer with high performance is selected to train the prediction model. The following will discuss the specific prediction process based on the actual collected data.

Using Windows 10 operating system, MATLAB R2016b and SQL Server 2000 database as software testing environment, using high-performance industrial computer as hardware testing platform. The parameter setting of neural networks in prediction model is realized by calling neural network toolbox.

The improved BP neural network algorithm is used to predict vital signs. With the increase of the number of neural network layers, the accuracy of the prediction results is also higher, but the network structure also becomes more complex, which leads to the longer training time of the neural network and the contradiction between the accuracy and efficiency of the neural network. Therefore, after many experiments, this paper uses three-layer BP neural network to establish the prediction model and obtains good test results.

To achieve faster convergence and improve learning efficiency, this paper chooses *traingdm*( ) function as the training function which can avoid the local minimum problem of neural network. In addition, starting with a relatively high learning rate (selecting 0.9 here), then slowly reducing the learning rate in the training, and finally determine the appropriate learning rate as 0.02. The activation function from the input layer to the hidden layer is *logsig*( ), which is a differentiable function that maps the input range of a neuron from (-∞,+∞) to (0,1) and the activation function from the hidden layer to the output layer is *purelin*( ), which is a linear function.

Before the experiment, the training samples and test samples of neural network are selected. In this paper, the data collected in the first 8 times are used to predict the ninth time. For example, the 9th data uses the 1st to 8th data to forecast, the 10th data uses the 2nd to 9th data to forecast, and so on. Here, this paper selects 20 groups of heart rate vital signs data of a subway driver as samples, which including 15 training samples and 5 test samples, as shown in Table 4 and table 5, respectively.

Table 4 Training samples (Times/Minute)

| Number | 3s | 6s | 9s | 12s | 15s | 18s | 21s | 24s | 27s | 30s |
|---|---|---|---|---|---|---|---|---|---|---|
| 1 | 79 | 80 | 80 | 81 | 82 | 80 | 81 | 80 | 81 | 81 |
| 2 | 96 | 98 | 98 | 97 | 95 | 91 | 90 | 90 | 89 | 90 |
| 3 | 93 | 92 | 88 | 86 | 83 | 79 | 75 | 72 | 73 | 73 |
| 4 | 80 | 84 | 85 | 85 | 84 | 83 | 83 | 82 | 81 | 82 |
| 5 | 89 | 88 | 88 | 91 | 92 | 92 | 89 | 87 | 84 | 86 |
| 6 | 79 | 81 | 83 | 84 | 86 | 86 | 84 | 82 | 78 | 79 |
| 7 | 97 | 95 | 94 | 95 | 95 | 94 | 94 | 93 | 92 | 91 |
| 8 | 86 | 87 | 87 | 86 | 85 | 82 | 81 | 79 | 79 | 80 |
| 9 | 84 | 85 | 86 | 86 | 87 | 87 | 87 | 85 | 82 | 83 |
| 10 | 93 | 95 | 96 | 98 | 97 | 96 | 95 | 94 | 92 | 93 |
| 11 | 90 | 93 | 95 | 93 | 90 | 87 | 84 | 81 | 77 | 80 |
| 12 | 83 | 81 | 81 | 80 | 79 | 80 | 81 | 83 | 84 | 83 |
| 13 | 87 | 90 | 91 | 90 | 89 | 88 | 86 | 85 | 85 | 86 |
| 14 | 84 | 86 | 88 | 89 | 90 | 89 | 88 | 86 | 86 | 88 |
| 15 | 94 | 95 | 96 | 98 | 97 | 95 | 94 | 94 | 93 | 93 |

Table 5 Test samples (Times/Minute)

| Number | 3s | 6s | 9s | 12s | 15s | 18s | 21s | 24s | 27s | 30s |
|---|---|---|---|---|---|---|---|---|---|---|
| 1 | 82 | 80 | 79 | 80 | 81 | 82 | 84 | 86 | 88 | 86 |
| 2 | 80 | 79 | 78 | 77 | 77 | 78 | 79 | 79 | 80 | 80 |
| 3 | 91 | 90 | 88 | 87 | 86 | 84 | 85 | 84 | 85 | 87 |
| 4 | 93 | 91 | 91 | 90 | 90 | 91 | 92 | 94 | 95 | 93 |
| 5 | 73 | 74 | 73 | 74 | 74 | 73 | 72 | 71 | 71 | 72 |

Using the selected 15 groups of training samples to train the learning model and using the test samples to verify the prediction results. The prediction results are shown in Table 6.

Table 6 Test results

|        | Test results 1 (27s) | | | Test results 2 (30s) | | |
| :---: | :---: | :---: | :---: | :---: | :---: | :---: |
| Number | Predictive value | Measured value | Error | Predictive value | Measured value | Error |
| 1 | 88.23 | 88 | 0.23 | 86.25 | 86 | 0.25 |
| 2 | 79.86 | 80 | -0.14 | 79.88 | 80 | -0.12 |
| 3 | 84.89 | 85 | -0.11 | 87.18 | 87 | -0.18 |
| 4 | 95.19 | 95 | 0.19 | 93.21 | 93 | 0.21 |
| 5 | 71.72 | 72 | -0.28 | 71.71 | 72 | -0.29 |

According to the analysis of the test results in Table 6, it can be found that the absolute error between the predicted value and the actual value of heart rate monitoring of metro drivers by using the prediction model in this paper is less than 0.5, and the prediction accuracy can meet the monitoring requirements of the vital signs of metro drivers.

6.3 Performance evaluation

Neural network algorithm has the disadvantage of slow learning speed, and the multi parameter vital signs monitoring system designed in this paper is an online real-time monitoring system, which requires high real-time monitoring. Therefore, whether the detection efficiency of neural network algorithm can meet the system requirements needs further analysis. Here, the heart rate of subway drivers is still selected as the test data.

The number of nodes in the input layer is 8, the number of nodes in the output layer is 1, the error rate meeting the accuracy requirements is 0.5%, and the maximum number of training is 10000. In the training process of neural network algorithm, the number of hidden layer nodes and error rate are counted, as shown in Table 7.

Table 7 Process of neuron selection

| Number | Number of hidden layer nodes | Error rate |
| :---: | :---: | :---: |
| 1 | 12 | 0.71% |
| 2 | 16 | 0.66% |
| 3 | 18 | 0.61% |
| 4 | 21 | 0.55% |
| 5 | 22 | 0.53% |
| 6 | 25 | 0.44% |

It can be seen from table 7 that the weight coefficient and threshold value of BP neural network algorithm are adjusted in the calculation process. After six iterations, the optimal solution meeting the requirements is obtained.

In order to illustrate the practicability of the system developed in this paper in multi parameter vital signs monitoring, the predicted value of vital signs in a period of time is randomly selected to compare with the actual value, and the statistical results are shown in Figure 4.

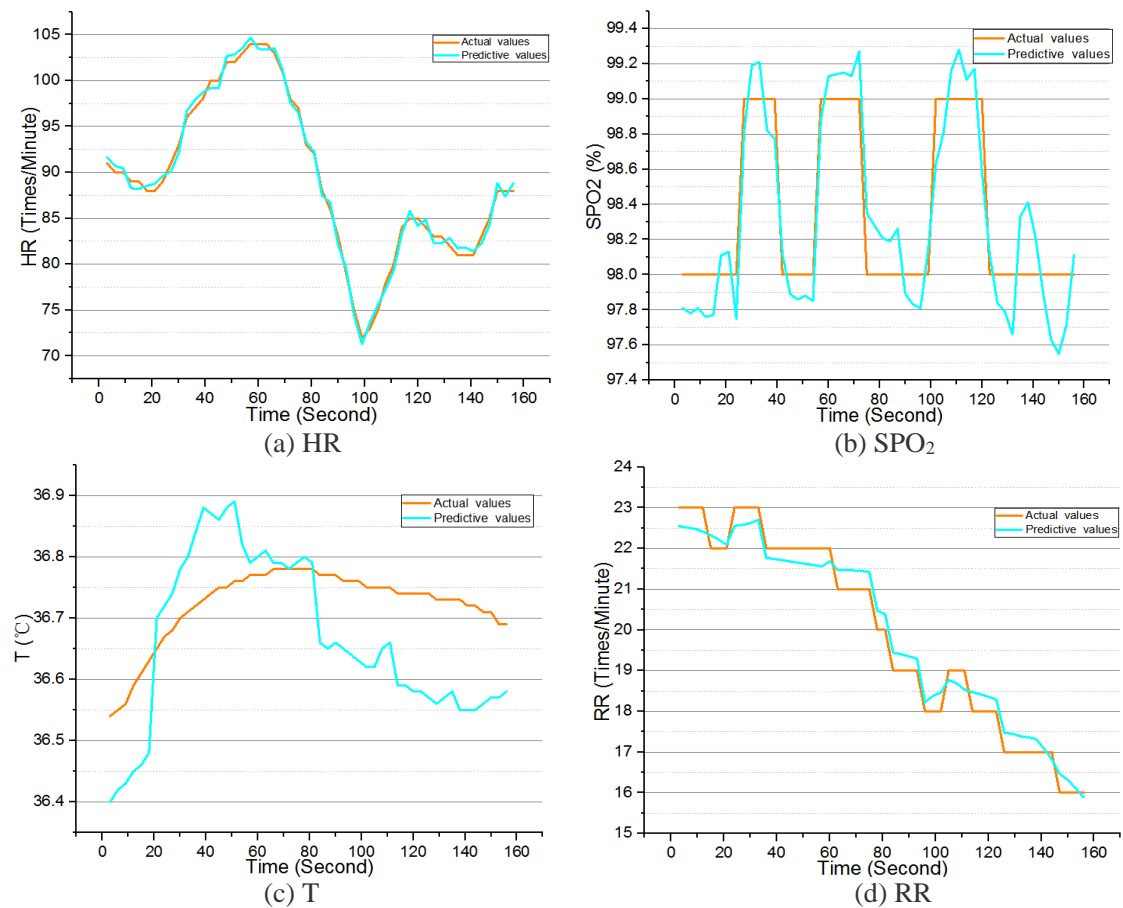

Figure 4 Predictive values and actual values

As can be seen from Figure 4, the deviation between the predicted value and the actual value of the system developed in this paper is not very large, which has a good prediction effect.

The prediction performance of the system developed in this paper needs to be further evaluated. TP rate and FP rate were used to evaluate the prediction results. Randomly select the data of 4 vital signs of 3 subway drivers for 4 consecutive hours, and the statistical results are shown in Table 8.

Table 8 Statistical results of FP rate and TP rate

| Subjects | Variables | FP rate | TP rate |
|---|---|---|---|
| 1 | $SPO_2$ | 1.54% | 95.68% |
| 1 | HR | 0.98% | 97.82% |
| 1 | RR | 1.26% | 96.97% |
| 1 | T | 0.77% | 98.56% |
| 2 | $SPO_2$ | 1.29% | 96.62% |
| 2 | HR | 1.12% | 97.03% |
| 2 | RR | 1.49% | 95.94% |
| 2 | T | 1.10% | 97.21% |
| 3 | $SPO_2$ | 1.07% | 97.28% |
| 3 | HR | 1.29% | 96.45% |
| 3 | RR | 0.89% | 98.37% |
| 3 | T | 1.51% | 95.83% |

It can be seen from the results in Table 8 that the system developed in this paper shows good accuracy and is very suitable for real-time monitoring of vital signs of subway drivers.

The risk assessment method of early warning score is adopted for the prediction results, and the assessment results are shown in Table 9.

Table 9 Results of risk assessment

| Subjects | Variables | Predictive value | Predictive value | Assessment score | Actual score |
|---|---|---|---|---|---|
| 1 | $SPO_2$ (%) | 97.33 | 97 | 0 | 0 |
| 1 | HR(Times/Minute) | 81.64 | 82 | 0 | 0 |
| 1 | RR(Times/Minute) | 17.83 | 18 | 0 | 0 |
| 1 | T (℃) | 35.89 | 36.20 | 1＊ | 0 |
| 2 | SPO$_2$ (%) | 90.51 | 91 | 2＊ | 1 |
| | HR(Times/Minute) | 88.75 | 89 | 0 | 0 |
| | RR(Times/Minute) | 17.32 | 17 | 0 | 0 |
| | T (℃) | 36.21 | 36.0 | 0 | 0 |
| 3 | SPO$_2$ (%) | 94.33 | 94 | 0 | 0 |
| | HR(Times/Minute) | 104.49 | 104 | 1＊ | 0 |
| | RR(Times/Minute) | 22.84 | 23 | 0 | 0 |
| | T (℃) | 35.81 | 35.6 | 1 | 1 |

Note: ＊ indicates miscalculation.

It can be seen from table 9 that there are misjudgments in the risk assessment of abnormal values of drivers' vital signs in the ground. Further analysis shows that misjudgment often occurs at the critical value of early warning range. This shows that although the system developed in this paper has a good monitoring effect, there is still a certain degree of conservatism.

# 7.Conclusion

In this paper, a real-time monitoring and early warning system for multi parameter vital signs of subway drivers is developed. Through the prediction and risk assessment of multiple vital sign parameters of subway drivers, the early warning information is sent to the dispatcher of the control center in advance to provide the necessary management basis for the safe operation of Metro. The sensors on the driver's head are used to collect a number of vital signs data, which are transmitted to the control center through wireless transmission technology, and the least mean square adaptive filtering algorithm is used to preprocess the data. An improved 3-layer BP neural network learning model is constructed, and the mathematical model of the network, the setting process of some important parameters in the model and the number of hidden layer nodes are introduced in detail. Selecting training samples for learning, and selecting test samples to verify the learning effect, led to an absolute error less than 0.5. The early warning score was used to evaluate the risk of abnormal vital signs, to provide the judgment basis for the dispatcher to make corresponding decisions. While the vital signs monitoring and early warning system of the subway driver under the condition of the train running at a constant speed is established in this paper, the operation condition of the train is complex and diverse. Therefore, the next key work is to study the real-time interaction of dynamic mechanism of various train operation conditions and their impacts on vital signs of drivers and passengers.

**Acknowledgments:** This study was partially supported by the LRail-SWJTU Consulting Program Program(R113620H01035) ,the State Key Laboratory of Rail Traffic Control and Safety Open Grant(RCS2017K008), and iTSS-RC/University of Waterloo, Canada.